\documentclass{rmaa}

\title{Centimeter emission in the UY Aur System}
\shorttitle{The UY Aur System}
\shortauthor{Contreras \& Wilkin}
\author{M.~E.~Contreras,\altaffilmark{1}
F.~P.~Wilkin\altaffilmark{1,2}
}

\altaffiltext{1}{Centro de Radioastronom\'{\i}a y Astrof\'{\i}sica, UNAM}
\altaffiltext{2}{Department of Physics and Astronomy, Union College, Schenectady, NY, USA}
\fulladdresses{
\item M.~E.~Contreras:
Centro de Radioastronom\'{\i}a y Astrof\'{\i}sica,
UNAM Campus Morelia, Apdo.Postal 3-72 (Xangari), 58089 Morelia, Michoac\'an,
M\'exico (m.contreras@astrosmo.unam.mx).
\item F.~P.~Wilkin:
Department of Physics and Astronomy, Union College, Schenectady, NY, 12308, USA 
(wilkinf@union.edu).
}

\resumen {
Reportamos observaciones de continuo a 3.6 cm tomadas con el Very Large Array (VLA) del
joven sistema binario UY Aur. La binaria est\'a compuesta por una estrella T Tauri, UY Aur A,
y una llamada ``Compa\~nera Infrarroja'' (IRC), UY Aur B separadas por 0$\rlap.{''}$89. 
UY Aur es un sistema interesante porque muestra caracter\'\i sticas observacionales cuyo origen
no se entiende del todo. Una de estas caracter\'\i sticas es el \'\i ndice espectral
inusualmente bajo que se encuentra en la regi\'on del milim\'etrico. En nuestro estudio con 
el VLA, hemos detectado radiaci\'on continua centim\'etrica que coincide con las posiciones 
medidas a 1.3 y 2.7 mm y es consistente con la posici\'on \'optica de UY Aur. Concluimos que la
emisi\'on a 3.6 cm est\'a asociada con el sistema binario. M\'as a\'un, sugerimos que la emisi\'on 
centim\'etrica podr\'\i a estar relacionada con un flujo bipolar previamente reportado.
}

\abstract {
We report 3.6 cm continuum observations taken with the Very Large Array (VLA) of the young
binary system UY Aur. The binary consists of a T Tauri star, UY Aur A, and a so-called 
``infrared companion'' (IRC),  UY Aur B, separated by $0\rlap.{''}89$. UY Aur is an interesting 
system because it shows observational features whose origin is not well understood. One of them 
is the unusual low spectral index found in the millimeter region. In our VLA study, we have 
detected centimeter continuum radiation that coincides with the reported positions at 1.3 and 
2.7 mm and is consistent with the optical position of UY Aur. We conclude that the 3.6 cm 
emission is associated with the binary system. Furthermore, we suggest that
the centimeter emission might be related to a previously reported bipolar outflow.}


\begin{document}
\maketitle


\section{Introduction}

The IRC system UY Aur was first reported as a visual double star 
by Joy \& van Biesbroeck (1944). Later, based on an infrared speckle study 
Ghez, Neugebauer \& Matthews (1993) and Leinert et al. (1993) confirmed that UY Aur 
is a binary system. Recently, Hartigan \& Kenyon (2003) reported the main properties 
of a sample of subarcsecond binaries in the Taurus-Auriga cloud based on HST spectra. They 
report UY Aur as a binary system composed of two classical T Tauri stars of spectral types 
M0 and M2.5 for the primary (UY Aur A) and the secondary (UY Aur B) respectively, separated 
by 0$\rlap.{''}89$ ($\sim$ 125 AU at 140 pc). 

  The system has been studied in detail in the infrared at $J, H \, \&  \,K'$ by Close
et al. (1998). Using infrared adaptive optics, Close et al. detected a circumbinary disk of 
$\sim$ 500 AU radius. In order to reproduce the spectral energy distribution of UY Aur A 
and B, they include in their models small inner disks around each star. The derived 
radii of the circumstellar disks are about 10 and 5 AU for components A and B, 
respectively. The Close et al. images also suggest that both inner disks are being fed by 
the outer circumbinary disk through
thin streamers of material. In the millimeter region,  both line emission ($^{13}$CO) 
as well as continuum at 2.7 and 1.3 mm were reported by Duvert et al. (1998). They have 
imaged the emission from the circumbinary disk in the $^{13}CO \;  \; J=1 \rightarrow 0$ and 
$J=2 \rightarrow 1$ transitions. Their spectral line observations agree well with
the infrared adaptive optics circumbinary disk reported by Close et al. not only in
position but in extent. Regarding the suggested small circumstellar disks, Duvert et al.
proposed that the 2.7 and 1.3 mm continuum emission can be attributed to partially resolved 
circumstellar disks around each star, with some possible contribution of free-free 
radiation. 

  In this work we report the first detection of centimetric emission at the position of 
UY Aur. We conclude that our 3.6 cm continuum detection is associated with the UY Aur system
and we discuss a possible origin of it.

\section{Observations}

\begin{figure}[!t]\centering
\begin{center}
\includegraphics[width=18pc]{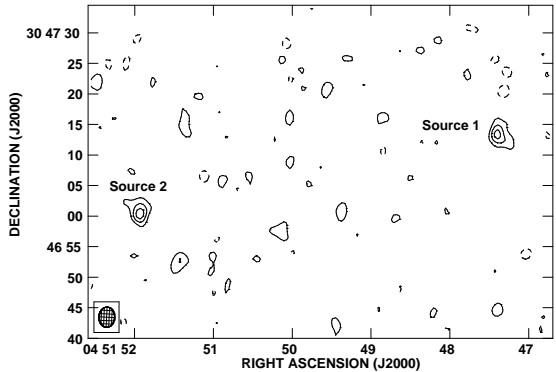}
\caption{CLEANed natural weight map at 3.6 cm. The map shows the two sources
detected in the UY Aur region. The peak of source 1 is located very close to the UY Aur
position and is suggested to be associated to the binary system. Contours are 
-2,2,4,6 times 16$\mu$Jy beam$^{-1}$}
\end{center}
\end{figure}

\begin{table}[!b]
\begin{center}
\small
\caption{Source Parameters}
\begin{tabular}{lccc}
\hline \\[-1ex]
Source & $\alpha$(2000) & $\delta$(2000) & $S_{3.6cm}$  \\
&h  m  s & $^{\circ}$  $'$  $''$ & [mJy] \\[2ex]
\hline \\[-2ex]
1 & 04 51 47.37 & 30 47 13.3 & 0.12$\pm$0.03 \\[0.7ex]
2 & 04 51 51.93 & 30 47 00.4 & 0.11$\pm$0.03 \\[2ex]
\hline
\end{tabular}
\end{center}
\hspace{0.1cm}
{{\scshape Note}.$-$ Absolute position errors are $\sim 0{\rlap.}{''}2$.}
\end{table}

Our 3.6 cm observations were made with the Very Large Array (VLA) of the 
NRAO\footnote{The National Radio Astronomy Observatory is operated by Associated 
Universities Inc. under cooperative agreement with the National Science Foundation}
on 2002 October 9th. The array was in the C configuration giving an angular resolution
of $\sim 2\rlap{.}{''}3$ and a total on-source integration time of $\sim 51$ minutes
was obtained. The amplitude and phase calibrators were 0137+331 and 0443+346, respectively. 
The bootstrapped flux density for 0443+346 was 0.615 $\pm 0.001$ Jy.

The data reduction was performed using the Astronomical Image Processing System (AIPS) software
of the NRAO. We have followed standard VLA procedures for editing, calibrating and
imaging. Figure 1 shows a natural weight CLEANed map of the UY Aur region. In this map two 
radio sources were detected at a 6-$\sigma$ level. We will refer to these sources as 
Sources 1 and 2. The peak of Source 1 is located very close to the UY Aur position. Flux 
densities and source positions were obtained using the AIPS IMFIT procedure. The apparent 
elongation of Source 1 is not real but it is due to beam deconvolution. Actually, the position 
angle of both Source 1 and the beam is the same: P.A.$= 177 ^\circ$. Besides, small structures 
present in Source 1 (Fig. 2) are not reliable since they are just at a 2$-\sigma$ level above 
rms-noise. Thus, since neither Source 1 nor Source 2 are spatially resolved, we have only 
determined integrated flux densities and source positions (see Table 1) from a 2D-Gaussian fit 
to each source.

\section{Discussion}

\begin{figure*}[!t]
\begin{center}
\includegraphics[width=27pc,height=37pc,angle=-90]{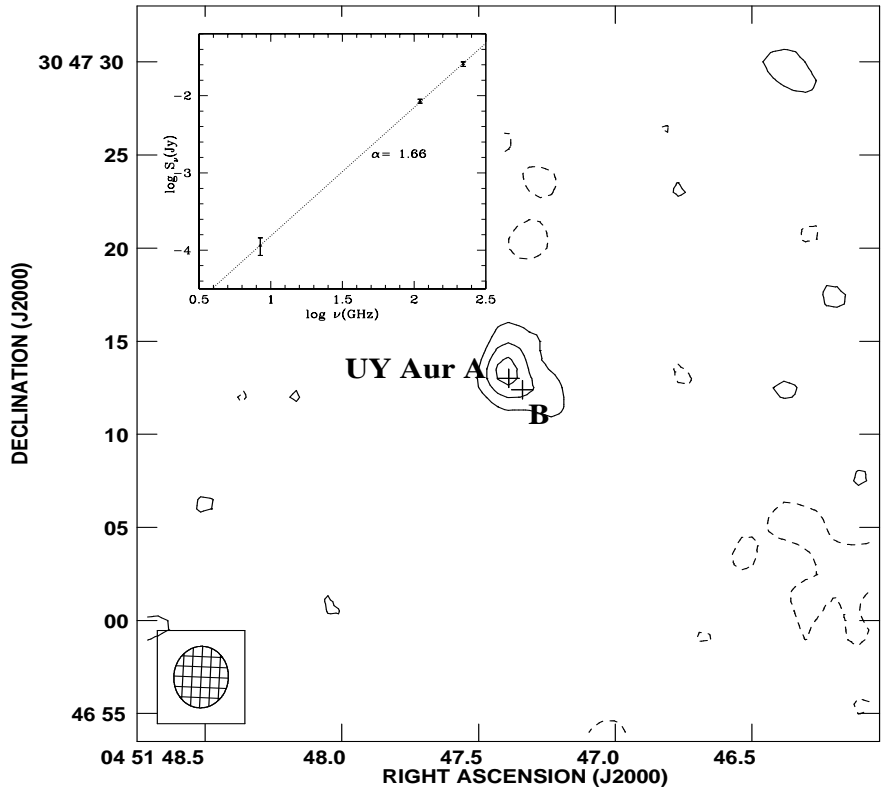}
\caption{This map shows an enlarged image of Fig. 1 around UY Aur. Crosses indicate 
the 
location of the UY Aur components corrected for precession and proper motion. The inset 
shows the least squares fit of the spectral index of the cm and mm emission. Contours as 
in Fig.1}
\end{center}
\end{figure*}

Both continuum and line emission are associated with the UY Aur system. Based 
on their continuum emission at 1.3 and 2.7 mm, Duvert et al. (1998) found an unusual
spectral index of 1.6$\pm$0.2, which is lower than the one expected for thermal dust emission,
$\sim$2.5 (Dutrey et al. 1996). Duvert et al. suggest that this low spectral index is the
result of a combination of two independent sources: one emitting free-free radiation
(showing a flat spectrum) and a second source showing normal dust emission (with a steep
positive index). They conclude that the free-free source should be located nearly coincident 
with or slightly north-east of the primary and propose centimetric observations to confirm 
this hypothesis.

In this work, we have observed the UY Aur system as part of a radio survey of IRCs. We detected
two 3.6 cm continuum sources, Sources 1 and 2 (see Fig. 1), at a 6$-\sigma$ level in the 
UY Aur region. On one hand, Source 2 does not have a known counterpart at any other wavelength.
The {\em a priori} probability of finding a 3.6 cm source with a flux density
of $\geq$0.11 mJy in a region of 2$'$ by 2$'$ is $\sim$ 0.16 (Windhorst et al. 1993). 
It is then quite likely that Source 2 is just a background source. On the other hand,
Source 1 is located very close to the reported  position of UY Aur 
(Herbig \& Bell 1988, hereafter HBC). Then, we propose that our centimeter source is related
to the binary system. In order to confirm that our detected emission is 
associated with it, we have precessed and corrected for proper motion the 
position of UY Aur according to Jones \& Herbig (1979). The resulting position for UY Aur 
coincides with that of our Source 1 to within $0\rlap.{''}34$, which according to 
Duvert et al. (1998) is less than the 1-2$\sigma$ uncertainty in the optical position and 
proper motion. Since the main component of the binary (UY Aur A) is the brightest
star in the system at optical and infrared wavelengths, we have assumed that the coordinates 
given in the HBC catalog belong to UY Aur A. Once the position of UY Aur A is fixed, 
we have derived the second component position relative to it (see Fig. 2) by taking a binary
separation of $0{\rlap.}{''}894$ and a position angle of $228.8^{\circ}$ (Brandeker,
Jayawardhana \& Najita 2003). On the one hand, our centimeter detection coincides with the 
position of UY Aur to within $0{\rlap.}{''}34$ and on the other hand it is consistent with the 
peak positions of the 1.3 and 2.7 mm emission reported by Duvert et al. to within $0\rlap.{''}2$.
Besides, although the low flux of our detection, it is consistent with the lowest centimetric 
emission of 0.1 mJy present in almost all outflow sources (Reipurth et al. 2004).
Therefore, we conclude that our detected 3.6 cm emission is associated with the UY Aur binary 
system. 

Regarding the spectral index, we have obtained a least squares fit to the millimeter
and centimeter fluxes. The resulting spectral index, $\alpha = 1.66$, is consistent with
that reported by Duvert et al. They have suggested that this low
value may be a combination of normal dust emission and free-free radiation from a stellar
wind or a jet. However, since both emissions follow a power law distribution ($\sim$2 for
dust radiation and 0.6 for a stellar wind), it is not possible to sum them and still obtain 
a single power law distribution over such a long range of wavelength without a significant bend. 
Thus, a satisfactory fit to the observations as the result of combining two such 
distributions could not be obtained. Although the 3.6 cm flux may originate in free-free
radiation, its low value clearly demonstrates that free-free emission is not contaminating the 
mm flux and does not explain the mm index. Then, the fact that the same low index is maintained 
over a large wavelength range might be fortuitous or might be entirely due to thermal dust 
emission. 

Duvert et al. (1998) proposed that the millimeter emission originates in the 
circumstellar region, actually in the small circumstellar disks, one around each component 
of the binary. The low spectral index in this region could be explained by circumstellar
flat-disk models of D'Alessio, Calvet \& Hartmann (2001) where they show that disks whose spectral
index is smaller than 2 are flat and the observed millimeter emission is due to optically thick 
and cold material.
But what about the origin of our centimeter emission? The centimeter emission might originate
in a stellar wind or an ionized jet as Duvert et al. (1998) have suggested. This last possibility 
might be supported by a kinematic study of Hirth et al. (1994). They deduce the 
existence of a bipolar, high velocity flow at a P.A.= 40$^{\circ}$ (P.A.= 220$^{\circ}$) 
associated with UY Aur. Then, our 3.6 cm emission may be related to this outflow. 
If the 3.6 cm flux is due to free-free emission, from our single 
wavelength observation it is not possible to distinguish between emission from a jet 
and a stellar wind. However, since our radio detection falls along the same power law distribution
obtained from the millimeter observations (see inset of Fig.~2), we cannot discard the 
possibility that it might be the long wavelength continuation of the thermal dust emission. 
Further observations at higher resolution and additional wavelengths are needed.

\section{Summary}

  We report for the first time VLA continuum emission at 3.6 cm associated
with the binary IRC system UY Aur. Surprisingly, our centimetric emission follows closely the 
low spectral index obtained in the millimeter region. 
 This low index might be explained by
a flat, optically thick and cold circumstellar disk, in this case one or both of the two small 
circumstellar disks. On the other hand, if the 3.6 cm emission is due to
free-free radiation, it may be related to the bipolar outflow reported by 
Hirth et al. (1994). However, from our single observation it is not possible to distinguish 
between free-free emission from a jet or a stellar wind. Radio centimeter observations at 
other wavelengths and/or with higher resolution are required to further clarify the origin of the 
radio continuum emission.

\acknowledgments
We thank Luis F. Rodr\'\i guez and Paola D'Alessio for their valuable comments on this work. 
We acknowledge financial support from DGAPA-PAPIIT and CONACyT-Ciencias B\'asicas. 
F.P.W. also was supported by the NSF International Researchers Fellowship Program.

\end{document}